\documentclass[prl,aps,amssymb,twocolumn,showpacs,floatfix]{revtex4}

\usepackage{graphicx}
\usepackage{dcolumn}
\usepackage{bm}

\begin{document}


\title{A Tunable Anomalous Hall Effect in a Non-Ferromagnetic System}

\author{J. Cumings$^{1,3}$, L. S. Moore$^{1}$, H. T. Chou$^{2}$, K. C. Ku$^{4}$, G. Xiang$^{4}$,
S. A. Crooker$^{5}$, N. Samarth$^{4}$, D. Goldhaber-Gordon$^{1}$}

\affiliation{Departments of $^{1}$Physics and $^{2}$Applied Physics,
Stanford University, Stanford, CA 94305; $^{3}$Department of
Materials Science and Engineering, University of Maryland, College
Park, MD 20742; $^{4}$Department of Physics, Pennsylvania State
University, University Park, PA 16802; $^{5}$National High Magnetic
Field Laboratory, Los Alamos, NM 87545}

\date{\today}

\begin{abstract}
We measure the low-field Hall resistivity of a magnetically-doped
two-dimensional electron gas as a function of temperature and
electrically-gated carrier density.  Comparing these results with
the carrier density extracted from Shubnikov-de Haas oscillations
reveals an excess Hall resistivity that increases with decreasing
temperature.  This excess Hall resistivity qualitatively tracks the
paramagnetic polarization of the sample, in analogy to the
ferromagnetic anomalous Hall effect. The data are consistent with
skew-scattering of carriers by disorder near the crossover to
localization.
\end{abstract}
\pacs{75.50.Pp, 71.70.Ej, 85.30.Tv}

\maketitle
The transverse, or Hall, resistivity is a direct measure of the sign
and concentration of charge carriers in most materials.  However,
other subtler electronic properties can produce ``anomalous"
corrections to the Hall resistivity.  Such anomalous Hall effects
(AHE) are well-known for correlated-electron systems such as
ferromagnetic metals \cite{hurd}, type-II superconductors
\cite{hagen}, weakly-localized conductors \cite{goldman}, and
Kondo-lattice materials \cite{brandt}\@. Although it has been known
for almost as long as the Hall effect itself, the AHE in
ferromagnetic materials remains the subject of contemporary debate
and has gained renewed interest due to its close theoretical
connection to the recently discovered spin-Hall effect and to spin
transport in general
\cite{dyakonov,hirsch,zhang,kato,chalzaviel,culcer}\@. Since this
AHE often persists above the ferromagnetic Curie temperature, an
analogous AHE should be observable in a purely paramagnetic system,
in which the charge carriers are spin polarized by an external
magnetic field \cite{chalzaviel}\@.  Unlike ferromagnetic metals
whose spin polarization is fixed by their chemistry \cite{lee},
paramagnetic semiconductors present the additional advantage that
their magnetic properties can be smoothly tuned by varying carrier
density, magnetic field, or temperature in a given sample. This
provides an ideal opportunity to clarify the mechanisms of the
AHE\@. Unfortunately, previous studies in narrow gap semiconductors
(e.g.\ n-InSb) revealed only a very weak AHE, despite large
$g$-factors and strong spin-orbit coupling \cite{chalzaviel},
conditions which should favor observing the AHE\@.  Diluted magnetic
semiconductors (DMS) have shown a clear AHE, but only in samples
that exhibit hole-mediated ferromagnetism \cite{ohno1}, with limited
opportunity for tuning with an electric field \cite{ohno2}\@. Bulk
crystals of n-type DMS  are not ferromagnetic and, despite their
extremely large spin splitting, have exhibited no AHE in previous
studies \cite{shapira}.

Here, we report the observation of a robust and tunable AHE in a
purely paramagnetic two-dimensional electron gas in a DMS quantum
well.  Surprisingly, the effect is much larger than in earlier
studies of paramagnetic semiconductors, despite the presence of a
large bandgap and--hence--a weak spin-orbit coupling.  We show that
the strength of the AHE is electrically tunable in this system,
shedding new light on the origins of this class of phenomena and
suggesting the possibility of gate-tunable spin transport in similar
structures. Finally, we identify a remarkably simple dependence of
our AHE on classical scattering in the regime of localization.

We have chosen to study the AHE in magnetically-doped
two-dimensional electron gases (M2DEGs) \cite{smorchkova,knobel}
derived from a II-VI DMS \cite{furdyna}\@. This choice is dictated
by several factors: (a) the carriers have an unusually large
paramagnetic susceptibility, resulting in a significant spin
polarization that enhances the strength of the AHE; (b) the 2D
nature and moderately high mobility (due to modulation doping) allow
independent measurement of the carrier density through Shubnikov-de
Haas (SdH) oscillations; (c) the paramagnetic susceptibility can be
tuned by varying the temperature; (d) at a fixed temperature, other
properties such as the carrier density, Fermi energy and resistivity
are tunable using the electric field from a gate electrode on the
sample surface.
\begin{figure}[tbp]
\includegraphics[width=.35\textwidth]{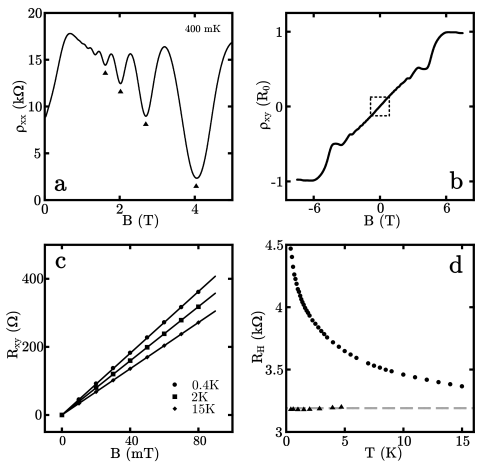}
\caption{a) \& b) The longitudinal and Hall resistances of a M2DEG,
revealing SdH oscillations and quantum Hall plateaus. The small box
in (b) schematically indicates the low-field region of (c)
($R_0=12.9 k\Omega$)\@. c) The Hall resistance at a variety of
temperatures. d) $R_H$ as a function of temperature.  The density
indicated by the arrows in (a) would be expected to give an $R_H$ of
3190~$\Omega$/Tesla, as indicated by the dashed line in (c)\@. The
triangles show the results of similar SdH measurements performed
over a range of temperatures.} \label{fig1}
\end{figure}

The samples consist of a modulation-doped single quantum well (10.5
nm thickness) of Zn$_{1-x-y}$Cd$_y$Mn$_x$Se ($x\sim$0.02,
$y\sim$0.12) sandwiched between ZnSe barriers. Symmetrically placed
n-type ZnSe layers (25 nm thick), spaced 12.5 nm from the quantum
well by intrinsic ZnSe spacer layers, donate free electrons to the
well. The material is described in more detail elsewhere
\cite{smorchkova,knobel}\@. The Mn ions behave essentially as free
spin-5/2 moments with Brillouin-like paramagnetic susceptibility.
Polarization of the Mn induces a spin-splitting in the conduction
electrons through an $s-d$ exchange coupling, giving the carriers an
effective $g$-factor far higher than even small-bandgap
semiconductors: $g\sim80$ in our material at 1.5 K, producing
complete polarization of carriers at $\sim$ 1 T\@.  The structures
were patterned into 100 $\mu$m wide Hall bars by photolithography
and wet-etching, and annealed indium metal provided ohmic contact to
the buried 2DEG\@.  In some samples, a gate electrode (10 nm Ti/100
nm Au ) was deposited by electron-beam evaporation. Transverse and
longitudinal resistances were measured in a $^3$He cryostat with a
base temperature of 290 mK using a 5 Hz lock-in technique at 30 nA
RMS excitation current \cite{resistance}.

In Figs.1 (a) and 1(b), we show the magnetic field dependence of the
longitudinal and transverse resistance, respectively, for a sample
at $T = $ 0.4 K, revealing SdH oscillations and an integer quantum
Hall effect at high fields.  The arrows marking the minima in Fig.\
1(a) correspond to the (spin-resolved) $\nu$=2, 3, 4, and 5 Landau
levels, consistent with a sheet density  $n_s = 1.96 \times 10^{11}$
cm$^{-2}$ which does not vary over the temperature range 0.4 K$< T
<$ 2 K. In Fig.\ 1(c), we show the field-dependence of the Hall
resistance ($R_{\rm{Hall}}$) at low fields. In this regime,
$R_{\rm{Hall}}$ is linear in $B$, with a slope usually identified as
the ordinary Hall coefficient, $R_H = (n_s e)^{-1}$\cite{Hall}\@.
However, $R_H$ clearly changes significantly between 0.4 K and 15 K,
even though there is no accompanying change in $n_s$ as determined
from SdH data. Figure 1(d) compares the temperature variation of
$R_H$ with that of the ordinary value calculated using $n_s$:
clearly, $R_H$ is anomalously high at low temperatures, gradually
approaching its ordinary value at higher temperatures.  As we show
later, the temperature-dependence of $R_H$ closely follows that of
the paramagnetic susceptibility of the magnetic ions. Similar
behavior is found upon a re-examination of low field Hall data from
measurements in other M2DEG samples \cite{knobel}, suggesting that
the phenomenon is generic to this system.

To tune the AHE, we fabricated a large-area gate electrode on a
second Hall bar from the same heterostructure.  Measurements of SdH
oscillations show that the application of a negative gate voltage
($V_g$) decreases $n_s$ (Fig.\ 2(a)) as expected.  Application of a
negative gate voltage also affects the measured value of $R_H$
(Fig.\ 2(b))\@. Consistent with a decrease in $n_s$, we find an
increase in $R_H$\@. In addition, $R_H$ becomes more strongly
temperature-dependent, indicating an increase in the strength of the
AHE\@.  To extract the AHE strength, we collapse all the data onto
the same curve by first subtracting the calculated value of the
ordinary Hall coefficient  and then dividing by a $V_g$-dependent
scale factor (deduced from a least-squares fit)\@.  Figure 2(c)
shows that this procedure collapses all data sets onto the same
characteristic T-dependence.  Figure 2(d) shows the extracted scale
factors, normalized to unity at $V_g = 0$, demonstrating that the
strength of the AHE can be electrically tuned by nearly a factor of
two.
\begin{figure}[tbp]
\includegraphics[width=.35\textwidth]{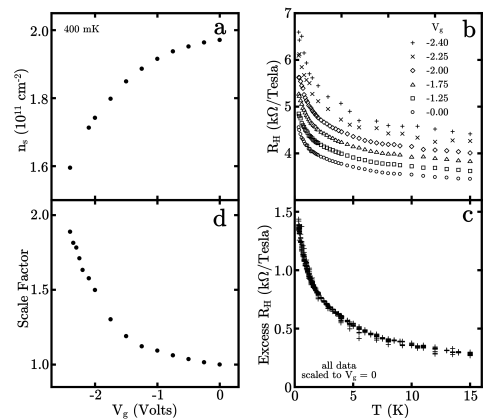}
\caption{The gate-voltage dependence of sample characteristics.  a)
Variation of carrier density (as determined from SdH
oscillations)\@. b) $R_H$ at various values of $T$ and $V_g$.  c)
All the data sets from (b) can be collapsed onto a single curve, by
subtracting the ordinary Hall coefficient for the density measured
in (a) and dividing by a scale factor.  d) The resulting scale
factors.} \label{fig3}
\end{figure}

The empirical form for the AHE is given by
\begin{equation}
R_{xy} = 1/(e n_s) B + R_s M.
\end{equation}
where the first term is the ordinary Hall resistance, proportional
to magnetic field $B$ and inversely proportional to the charge per
carrier ($e$) and the sheet density of carriers ($n_s$)\@. The
second term is the anomalous contribution, proportional to the
magnetic moment $M$ of the system. In the M2DEG studied here, $M$ is
the spin-polarization of the carriers, which is in turn proportional
to the magnetization of the local Mn moments at low fields.  The
magnetization of the paramagnetic local Mn moments is empirically
known to follow a modified Curie-Weiss law, $M \sim$ B/(T+T$_0$), in
the low field limit where T$_0$ is a phenomenological parameter that
accounts for the short-range coupling of neighboring Mn spins
\cite{furdyna}\@.  Eq.\ (1) thus predicts that $R_{xy}$ is linear in
$B$, and the AHE must be separated from the ordinary
temperature-independent Hall effect by examining the $T$ dependence
of $R_H$.

The AHE is commonly characterized by measuring $M$ and extracting
the constant of proportionality $R_s$ from Eq.\ (1)\@. In our
sample, $M$ is too small to measure by conventional susceptometry
methods and micro-fabricated cantilever measurements require
specially-designed heterostructures \cite{harris}\@. Hence, we
extract $M$ from the enhanced Zeeman shift of spin-sensitive
photoluminescence (PL) \cite{PL}\@.  In these measurements,
photo-excited electrons and holes form excitons, which then decay
radiatively.  The large Zeeman-shift of the PL peak is directly
proportional to the Zeeman-splitting and the relative magnetization
of the electrons in the conduction band, which accurately tracks the
Brillouin-like magnetization of the paramagnetic Mn local moments in
the M2DEG \cite{furdyna}\@.  To closely match the results of the
different measurements, neighboring pieces of the same sample were
used for PL and transport.
\begin{figure}[tbp]
\includegraphics[width=.35\textwidth]{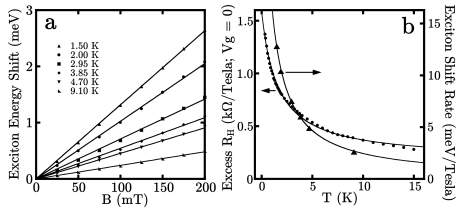}
\caption{Results of photoluminescence studies.  a) The exciton shift
as a function of $B$ for a variety of temperatures.  This shift is
proportional to electronic magnetization.  b) The low-field linear
coefficient of the exciton shift, superimposed on the low-field
excess $R_H$ (at $V_g$ = 0)\@.  The lines represent least-squares
fits as described in the text.} \label{fig4}
\end{figure}

In Fig.\ 3(a), we show results of the PL measurements, with fits to
a spin-5/2 Brillouin function, including weak antiferromagnetic
coupling characterized by $T_0$ \cite{furdyna}\@. These fits yield
$T_0$ ranging from 0.69 K to 1.03 K, in agreement with previous
studies of similar M2DEGs. In Fig.\ 3(b), we compare the low-field
behavior of the PL shift with that of the AHE by plotting the
T-dependence of both the linear coefficient of the PL shift and the
excess $R_H$ (at $V_g$ = 0)\@. The PL shift rate is fit to a
Curie-Weiss law $[C/(T+T_0)]$ and $R_H$ is fit to $[A +
C/(T+T_0)]$\@. These fits yield $T_0^{\rm{PL}} = 0.53$ and
$T_0^{\rm{Hall}} = 1.58$\@. The $A$ parameter of the $R_H$ fit gives
$n_s = 1.87 \times 10^{11}$ cm$^{-2}$ in good agreement with the
SdH-determined value of $1.96 \times 10^{11}$ cm$^{-2}$
\cite{Aparameter}\@. This agreement demonstrates that the excess
$R_H$ observed in these samples scales directly with the sample
magnetization and therefore shares common origins with the AHE of
ferromagnetic systems \cite{2Dscreen}.

We now turn to a discussion of theoretical models of the AHE and
their relation to our system.  We note at the outset that the
spin-orbit (SO) coupling in our system is relatively weak,
characterized by a parameter $\lambda$ more than 10 times smaller
than for GaAs and more than 1000 times smaller than for InSb.  Weak
SO coupling is confirmed by the absence of weak antilocalization in
low field magnetoresistance measurements of non-magnetic versions of
our host material \cite{smorchkova}\@. This low SO coupling demands
that we look in detail at the origin of the observed AHE\@.  Our
material could exhibit an AHE of intrinsic origin (unrelated to
disorder potentials) if there were an electric field perpendicular
to the 2DEG strong enough to induce a Rashba SO splitting
\cite{culcer}\@.  However, our quantum well is designed to be
symmetric, without any electric field. As an extreme case, if we
assume a completely asymmetric well, with the sheet density entirely
compensated by an electric field on one side, the strength of this
field would be $4 \times 10^6$ V/m.  Even in this limit--and with
all other assumptions as generous as possible--the intrinsic effect
could account for only 6\% of the observed deviation (at $V_g$ = 0,
T = 1K) \cite{culcer}\@.  Another potential source of intrinsic AHE
is the strain-induced SO splitting due to lattice mismatch at the
growth interface.  However, the leading contribution is from
off-diagonal strain due to shear stress, and a simple calculation
shows even this effect to be negligible. Therefore, intrinsic
effects are insufficient to account for our AHE, which must instead
be caused by disorder-induced scattering. A simple theory
incorporating both skew-scattering and sidejump-scattering predicts
an excess Hall resistivity:
\begin{equation}
\Delta\rho_{xy} = 2\pi Vm^*\lambda \langle\mu_z\rangle n_s \rho_{xx}
+ 2e^2 \lambda\langle\mu_z\rangle n_s \rho^2_{xx},
\end{equation}
where $\lambda$ is a measure of the strength of the SO coupling in
the material; $V$ is the potential of individual  $\delta$-function
scatterers, which can be either positive (repulsive) or negative
(attractive); $\langle\mu_z\rangle$ is the averaged spin magnetic
moment per carrier (in units of Bohr magnetons); and $\rho_{xx}$,
$m^*$, and $e$ are the longitudinal resistivity, effective mass, and
charge per carrier, respectively \cite{bernevig, nozieres}.\  The
first and second terms originate from skew scattering and sidejump
scattering, respectively.  In the II-VI quantum wells studied here,
$\lambda$ is known to be negative, as the $g$-factor for electrons
in the conduction band is reduced relative to its vacuum value due
to spin-orbit coupling in the semiconductor \cite{willatzen}.\  The
sidejump term hence has a sign opposite to that of the ordinary Hall
effect and cannot account for our observations. Further, the
strength of the sidejump term predicted by Eq. (2) only amounts to
about 1\% of the observed AHE (at $V_g$ = 0, T = 1K)
\cite{nozieres}.\  Having exhausted other options, we tentatively
link our positive AHE to skew scattering from scatterers with
attractive potentials (V negative).  This would be confirmed by
observing a clear linear dependence of the AHE on $\rho_{xx}$.

Figure 4(a) shows $\rho_{xx}$ as a function of both $V_g$ and $T$.
At the lowest temperatures, $\rho_{xx}$ varies by about a factor of
twenty as a function of $V_g$.  Within the same V$_g$ range, the
strength of the AHE (Fig. 2(d)) varies by only a factor of two,
clearly demonstrating a weaker-than-linear dependence on
$\rho_{xx}$.\ We are unaware of any previous theoretical prediction
of such weak $\rho_{xx}$-dependence for an AHE, and we suggest
localization as a plausible explanation.  In our sample, as the
strength of the gate voltage is increased $\rho_{xx}$ passes through
the quantum of resistance, $\sim$12.9 k$\Omega$, at about $V_g =
-1.5$ V and increases rapidly at low T, indicative of localization.
To account for this, we suggest the ansatz $\rho_{xx}(T) =
\rho_{\rm{classical}} + \rho_{\rm{localization}}(T)$, and propose
that $\rho_{\rm{classical}}$ replaces $\rho_{xx}$ in Eq. 2 for this
regime. To extract $\rho_{\rm{classical}}$ from our data, we use the
theory of two-dimensional variable range hopping (2DVRH), which
predicts \cite{brenig}
\begin{equation}
\rho_{xx}(T)=\rho_0 exp[(T_{VRH}/T)^{1/3}]
\end{equation}
This form yields an easily extractable high-temperature limit,
$\rho_0$.  To test if this is the correct form, in Fig. 4(a) we plot
$\rho_{xx}$ vs. T$^{-1/3}$ on a log scale which should give straight
lines for 2DVRH.  The data are not strictly linear, probably because
they are taken at the onset of localization and not in the regime of
strong localization.  Nevertheless, extrapolating the data from the
two highest temperature points yields $\rho_0$ values that we use as
estimates for $\rho_{\rm{classical}}$.  Figure 4(b) shows the AHE
scale factors from Fig. 2(d) plotted as a function of
$\rho_{\rm{classical}}$, in comparison with Eq. (2) \cite{Vg}.\  A
linear fit through the data extrapolates to the origin (with no
offset), demonstrating that the AHE strength scales linearly with
the Drude/Boltzmann classical resistance and offering further
evidence for the skew-scattering origin of the AHE.  If we use Eq.
(2) and the slope from Fig. 4(b), and we take the size of the skew
scattering sites to be $\lambda_F^2$, we extract a scattering
potential $V$ on the order of -20 meV, which is plausible for this
$\sim$2.5 eV gap semiconductor.
\begin{figure}[tbp]
\includegraphics[width=.35\textwidth]{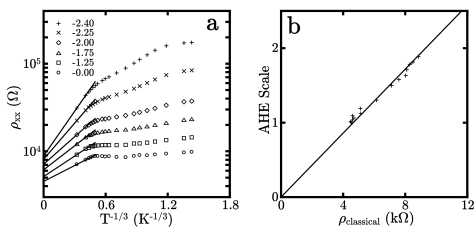}
\caption{The dependence of the AHE on sample resistivity.  a)
Longitudinal resistivity is plotted against T$^{-1/3}$ for a variety
of V$_g$ values.  The lines are extrapolations to
T$\rightarrow\infty$ from the two highest temperature points.  b)
The strength of the AHE (as measured by the scale factor in Fig.
2(d)) is plotted against the classical resistivity (as described in
the text). The line represents a single-parameter fit to the data
using a linear form with no constant term.} \label{fig5}
\end{figure}

In summary, we have reported the surprising observation of a robust
AHE in a non-ferromagnetic 2DEG, despite a very weak SO coupling.
Electrical gating allows us to study the AHE as a function of
carrier density, and--hence--disorder. Our data are consistent with
an AHE that originates in impurity-related skew scattering, but we
find clear deviations from standard theoretical expectations with
increasing disorder, particularly beyond the crossover to
localization \cite{burkov}.\  These results suggest the possible
emergence of new physics from the interplay between disorder and the
AHE, which we hope will motivate the development of new theories
that address this issue.  We finally note that, while the AHE
produces an excess Hall resistance in our M2DEG at low field, the
quantum Hall effect plateaus remain properly quantized at high
field.  This raises interesting questions about the interplay
between the two phenomena.

The authors thank A. H. MacDonald for initially suggesting the
measurements and B. A. Bernevig for extensive discussions.  The work
at Stanford was supported by the ONR under contracts
N00014-02-1-0986 and N00014-01-1-0569(YIP), and was performed in
part at the Stanford Nanofabrication Facility of NNIN supported by
the National Science Foundation under Grant ECS-9731293.  Work at
PSU was supported by ONR under contract N00014-02-1-0996.


\newpage


\begin{references}
\bibitem{resistance} Measured resistance values show no significant change when the current is
lowered to 10~nA\@.  Measurements at 150~nA show subtle differences
attributable to electron heating. The AC lock-in frequency of 5 Hz
was chosen to maximize measurement speed.  A phase lag in the
sample's response introduces a small systematic error in our
measurement of the Hall resistance.  This error is 0.1\% at 5 Hz,
and increases with increasing frequency.  As observed previously in
similar samples, the capacitance responsible for the phase lag is
associated with the heterostructure, not the measurement setup.
\bibitem{Hall} $R_H$ is determined by fitting the R$_{xy}$ data to a linear form in the range of 90
mT $< B <$ 90 mT\@.  For some pairs of sample contacts, a strong
capacitive signal of unknown origin is observed below the critical
field of the superconducting indium contacts (H$_c\sim$ 25~mT)\@. In
these cases, the fitting is confined to the range 30 mT $<|B|<$ 90
mT\@.  Polynomial fits were also attempted up to cubic or quintic
terms, without significant change to the extracted $R_H$ values.
Analyzing the Hall conductivity, $\sigma_{xy}$, instead of R$_{xy}$
results in no change to the interpretation of the AHE.
\bibitem{Aparameter} Fixing the $A$ parameter to exactly match the SdH density gives a noticeably worse fit and a
$T_0^{\rm{Hall}} = 2.93$\@.  $T_0$ is a phenomenological parameter
that depends sensitively on the details of the fit; the difference
between $T_0^{\rm{PL}} = 0.53$ and $T_0^{\rm{Hall}} = 1.58$ is not
alarming.
\bibitem{Vg} The leading gate-voltage dependence of $\Delta\rho_{xy}$ from the first term of Eq.\ (2) is $\rho_{xx}$\@.  The mean
carrier magnetic moment $\langle\mu_z\rangle$ is proportional to
$n_s^{-1}$ in a simple Fermi gas model, and the quantity
$\langle\mu_z\rangle n_s$ therefore has no dependence on density or
gate-voltage.  All other parameters are expected to be constant.
\end{references}
\end{document}